\documentclass[aps,prd,reprint,groupedaddress,showpacs]{revtex4-1}
\usepackage{amssymb}
\usepackage{graphicx}
\usepackage{dcolumn}
\usepackage{amssymb}

\begin{document}

\title{Dynamics of the inflaton scalar field for a certain class of E-model potentials}
\author{Vladimir A. Koutvitsky}
\author{Eugene M. Maslov}
\email{eugen.masloff2014@yandex.ru}
\affiliation{Pushkov Institute of Terrestrial Magnetism, Ionosphere and Radio Wave Propagation (IZMIRAN) 
of the Russian Academy of Sciences,\\ Moscow, Troitsk, Kaluzhskoe Hwy 4, Russian Federation, 108840}
\date{\today}

\begin{abstract}
We investigate the dynamics of the inflaton scalar field in a certain class
of inflationary E-models that combine the properties of the Starobinsky
model and the $\alpha $-attractor model. The inflaton potential we are
dealing with has an exponentially flat plateau at high field values and a
sharply defined minimum at zero. Using the slow-roll approximation, we
obtain the analytic expressions describing evolution of the background
inflaton field at the inflationary stage. To describe the nonlinear field
oscillations at the preheating stage, we use the technique of separation of
fast (oscillation phase) and slow ( field energy density) variables. The
obtained expressions are in good agreement with the results of direct
numerical integration of the field equations. These expressions are then
used to study the evolution of perturbations at the preheating stage. Based
on the Mukhanov-Sasaki equation, we derive the Hill equation with slowly
varying parameters, which describes the scalar perturbation modes taking
into account the anharmonicity of the background oscillations. Using Floquet
theory, we analyze the structure of the resonance zones of this equation and
then integrate it numerically. We show that the cosmological expansion
limits the resonant growth of scalar modes, making their amplitudes nearly
constant at late times. We also derive the corresponding Hill equation for
tensor modes. We show that resonant amplification of tensor fluctuations of
the metric does not occur.\end{abstract}
\maketitle

\section{Introduction}

The currently generally accepted hypothesis is that the evolution of the
Universe began with exponential expansion at the so-called inflationary
stage. This hypothesis allows us to explain in a completely natural way the
flatness, homogeneity and isotropy of the modern Universe on large scales,
as well as to interpret the tiny anisotropy in the cosmic microwave
background observed by the WMAP and Planck satellites and to suggest the existence of
relict gravitational waves. \cite{STAROBINSKY197930, STAROBINSKY198099, PhysRevD.23.347, 
LINDE1982389, PhysRevLett.48.1220}. It is believed that this expansion was caused 
by the existence of a fundamental scalar field $\phi (t)$. The dynamics of such a field in a
homogeneous and isotropic Universe with the Friedmann-Robertson-Walker
metric is described by the equation%
\begin{equation}
\phi _{tt}+3H\phi _{t}+V^{\prime }(\phi )=0,  \label{eq1}
\end{equation}%
where $a(t)$ is the scale factor, $H=a_{t}/a$ is the Hubble parameter, $%
V(\phi )$ is the potential of the inflaton scalar field. For the Universe
filled with such a field, the Friedmann-Raychaudhuri equations have the form%
\begin{equation}
H^{2}=\frac{8\pi G}{3}\rho ,\qquad \frac{a_{tt}}{a}=-\frac{4\pi G}{3}(\rho
+3p),  \label{eq2}
\end{equation}%
where the energy density $\rho $ and the effective pressure $p$ are%
\begin{equation}
\rho =\phi _{t}^{2}/2+V(\phi ),\qquad p=\phi _{t}^{2}/2-V(\phi ).
\label{eq3}
\end{equation}%
It follows that%
\begin{equation}
\rho _{t}=-3H(\rho +p),\qquad H_{t}=-4\pi G\phi _{t}^{2}.  \label{eq4}
\end{equation}%
If the potential $V(\phi )$ has a sufficiently flat region in a certain
interval of $\phi $ values, then the rate at which the field $\phi $ rolls
towards a global minimum of the potential will be small within this interval,
and, according to (\ref{eq4}), $H\approx const$, which leads to the
exponential expansion.

After the end of the slow-roll stage the inflaton field begins to oscillate
rapidly around the minimum of the potential, finally decaying into
radiation. This phase is usually called reheating. However, there is a short
period of reheating, immediately after the end of inflation, when the nearly
uniform inflaton field oscillates coherently. At this stage, named as
preheating \cite{PhysRevD.56.3258}, perturbations of the inflaton field can grow
exponentially due to parametric resonance, leading to subsequent decay of
the field (see, e.g., \cite{RevModPhys.78.537} for a review).

To date, several dozen different potentials have been proposed to implement
these scenarios \cite{MARTIN201475}. Such potentials, which are in good agreement
with modern observational data \cite{Planck2018-X}, include, in particular, the
Starobinsky potential, conformally related to the $R+R^{2}$ inflationary
model \cite{STAROBINSKY198099}. It has the form%
\begin{equation}
V(\phi )=V_{\infty }\left( 1-e^{-b\phi }\right) ^{2},  \label{eq5}
\end{equation}%
where $b=\sqrt{2/3}M_{Pl}^{-1}=\sqrt{16\pi G/3}$, \ $V_{\infty
}=m^{2}/\left( 2b^{2}\right) $ is the value of the potential at infinity, $m$
is the inflaton mass. Another example is the potential of superconformal 
$\alpha$-attractor model \cite{Kallosh2013},%
\begin{equation}
V(\phi )=V_{\infty }(1-e^{-\alpha ^{-1/2}b\phi })^{2},  \label{eq6}
\end{equation}%
where $\alpha $ is a free parameter. Both of them belong to the class of
potentials that have an exponentially flat plateau, the so-called E-model
potentials.

In the present paper we study the dynamics of the inflaton scalar field in
the E-model with potential of the form%
\begin{equation}
V(\phi )=V_{\infty }\left( 1-e^{-f(\chi )}\right) ^{2},  \label{eq7}
\end{equation}%
where $\chi =b\phi $, and $f\left( \chi \right) $ is an increasing odd
function with asymptotics%
\begin{equation}
f\left( \chi \right) \thicksim C\chi \quad (C>1,\text{ }\chi \rightarrow
0),\quad f\left( \chi \right) \thicksim \chi \quad (\chi \rightarrow \pm
\infty ).  \label{eq8}
\end{equation}%
Thus, potential (\ref{eq7}) behaves like the Starobinsky potential for large
values of $\chi $, and like the $\alpha$-attractor potential for small values
of $\chi $.

Our paper is organized as follows. In Sec. II, having chosen a specific
function $f\left( \chi \right) $ with asymptotics (\ref{eq8}), we
investigate the dynamics of the background inflaton field. We consider the
slow-roll stage, the preheating stage, and compare the obtained analytical
solutions with the results of numerical integration. In Sec. III we consider
the evolution of perturbations arising at the preheating stage. First, we
study perturbations of the inflaton background. Then, we examine the
influence of scalar metric perturbations on them. Tensor perturbations are
also considered. Some remarks regarding the obtained results are made in
Sec. IV.

\section{Dynamics of the background inflaton field}

\begin{figure}[ht]
\begin{center}
\includegraphics[width=0.45\textwidth]{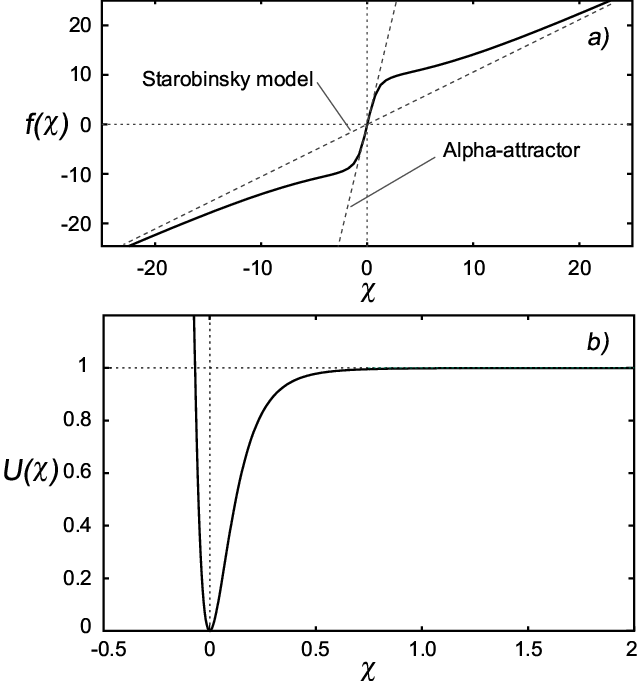}
\end{center}
\caption{Characteristic form of the function $f(\chi)$ (a) and the corresponding potential (b)\label{Fig1}}
\end{figure}

In dimensionless variables $\chi =b\phi ,$ $\tau =bV_{\infty }^{1/2}t$ Eq. (%
\ref{eq1}) with potential (\ref{eq7}) takes the form%
\begin{equation}
\chi _{\tau \tau }+3\mathcal{H}\chi _{\tau }+U_{\chi }=0,  \label{eq9}
\end{equation}%
where%
\begin{equation}
\mathcal{H}=a_{\tau }/a=\left( \varrho /2\right) ^{1/2},  \label{eq10}
\end{equation}%
\begin{equation}
\varrho =\rho /V_{\infty }=\chi _{\tau }^{2}/2+U(\chi ),  \label{eq11}
\end{equation}%
\begin{equation}
U(\chi )=\left( 1-e^{-f(\chi )}\right) ^{2}.  \label{eq12}
\end{equation}%
In what follows we will deal with the potential (\ref{eq12}), where the
function $f(\chi )$ is given by the expression%
\begin{equation}
f(\chi )=\chi \left( 1+\frac{\alpha }{\left( \beta ^{2}+\chi ^{2}\right) ^{p}%
}\right) ^{q}  \label{eq13}
\end{equation}%
with parameters $\alpha\! >0$,\!\! \ $\beta\! >0$,\!\! \ $1\!<2p\leqslant\! 1/q$. (see Fig. \ref{Fig1}). 

\subsection{Slow-roll stage}

\begin{figure}[ht]
\includegraphics[width=0.42\textwidth]{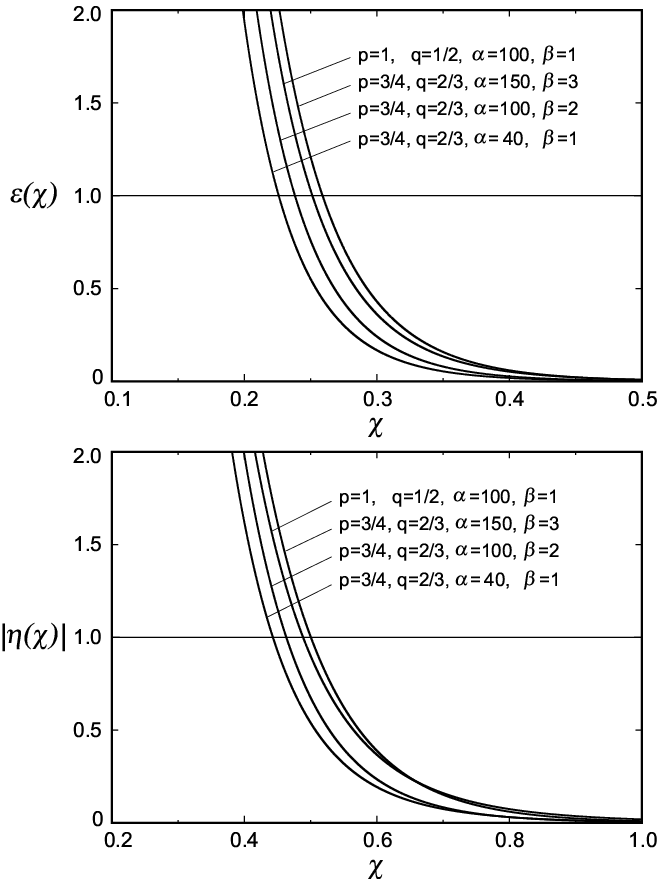}
\caption{The slow-roll parameters $\varepsilon(\chi) $ and $\eta(\chi) $ for potential (\ref{eq12})\label{Fig2}}
\end{figure}

For the potential (\ref{eq12}), there is a range of $\chi $ values in which%
\begin{equation}
\varepsilon =\frac{1}{3}\left( \frac{U_{\chi }}{U}\right) ^{2}\ll 1,\quad
\left\vert \eta \right\vert =\frac{2}{3}\left\vert \frac{U_{\chi \chi }}{U}%
\right\vert \ll 1.  \label{eq14}
\end{equation}%
In this region, the second-derivative term in equation (\ref{eq9}) can be
neglected, so that the field $\chi $ will slowly roll down from some initial
value $\chi _{ini}$ to the minimum of the potential according to the
equation%
\begin{equation}
\chi _{\tau }=-\frac{U_{\chi }}{3\mathcal{H}}.  \label{eq15}
\end{equation}%
In this regime%
\begin{equation}
\frac{\chi _{\tau }^{2}/2}{U(\chi )}\approx \frac{\varepsilon }{3}\ll
1,\quad \mathcal{H}\approx \left( \frac{1}{2}U(\chi )\right) ^{1/2}\approx
const,  \label{eq16}
\end{equation}%
so that $a\sim e^{\mathcal{H}\tau }$. Calculation gives%
\begin{equation}
\varepsilon =\frac{4}{3}\left( \frac{D_{1}}{e^{f}-1}\right) ^{2},
\label{eq17}
\end{equation}%
\begin{equation}
\eta =\frac{4}{3\left( e^{f}-1\right) }\left[ D_{2}-\left( 1-\frac{1}{e^{f}-1%
}\right) D_{1}^{2}\right] ,  \label{eq18}
\end{equation}%
where%
\begin{equation}
D_{1}=\left( 1+\frac{\alpha }{\left( \beta ^{2}+\chi ^{2}\right) ^{p}}%
\right) ^{q-1}\left[ 1+\alpha \frac{\beta ^{2}+\left( 1-2pq\right) \chi ^{2}%
}{\left( \beta ^{2}+\chi ^{2}\right) ^{p+1}}\right] ,  \label{eq19}
\end{equation}%
\begin{eqnarray}
\hspace{-4mm}D_{2} &=&\frac{2\alpha p\,q\chi }{\left( \beta ^{2}+\chi ^{2}\right) ^{p+2}}%
\left( 1+\frac{\alpha }{\left( \beta ^{2}+\chi ^{2}\right) ^{p}}\right)
^{q-2}  \nonumber \\
&\times&\! \left[\left( 2p\!-\!1\right) \chi^{2}\!-\!3\beta ^{2}\!-\!\alpha \frac{
\left( 1\!-\!2p\,q\right) \chi ^{2}+3\beta ^{2} }{\left( \beta ^{2}+\chi^{2}\right) ^{p}}\right]\! .  \label{eq20}
\end{eqnarray}%
Fig. \ref{Fig2} shows the dependences $\varepsilon \left( \chi \right) $ and $%
\left\vert \eta \left( \chi \right) \right\vert $ for several values of the
parameters involved in the function $f\left( \chi \right) $. 

The slow-roll stage ends when the field $\chi $ reaches the value $\chi_{end}$ determined  from the condition%
\begin{equation}
\max \{\varepsilon (\chi ),\left\vert \eta (\chi )\right\vert \}=1.
\label{eq21}
\end{equation}%
Thus, from some moment $\tau $ until the end of inflation, the Universe
expands by $e^{N_{sr}}$ times, where%
\begin{equation}
N_{sr}(\chi )=\ln \frac{a(\tau _{end})}{a(\tau )}\approx \frac{\sqrt{3}}{2}%
\int_{\chi _{end}}^{\chi }\frac{d\chi }{\varepsilon ^{1/2}(\chi )}.
\label{eq22}
\end{equation}%
The total number of e-foldings from the beginning of inflation is believed
to be $N_{sr}(\chi _{ini})\gtrsim 60$ (see, e.g., \cite{Gorbunov}) that
estimates the value $\chi _{ini}$ by Eq. (\ref{eq22}).

As is known, the parameters $\varepsilon $ and $\eta $ make it possible to
evaluate the observed spectral characteristics of the perturbations
generated at the slow-roll stage using the formulas

\begin{tabbing}
\hspace{2mm} \= \hspace{45mm} \= \hspace{45mm}\= \kill

\> $n_{s}\simeq 1-6\varepsilon \left(\chi _{\ast }\right)+2\eta \left( \chi_{\ast }\right)$ \> (scalar spectral index),\\
\> $n_{t}\simeq -2\varepsilon \left( \chi _{\ast }\right)$ \> (tensor spectral index), \\
\> $r\ \simeq 16\varepsilon \left( \chi _{\ast }\right) 		$ \> (tensor-to-scalar ratio),  
\end{tabbing}
where $\chi _{\ast }$ denotes the values of the background inflaton field
at which physically interesting perturbation modes cross the horizon. It is
assumed that this occurred in the range of 50$\div $60 e-foldings before the
end of inflation, which allows us to estimate $\chi _{\ast }$ using Eq. (\ref%
{eq22}). Table 1 presents the spectral indices calculated at $N_{sr}(\chi_{\ast })=55$ 
 for some values of the parameters in the function 
$f(\chi) $ (\ref{eq13}). It should be noted that the Planck 2018
observations yield $n_{s}\!=\!0.9649\pm\! 0.0042$, $n_{t}\!<\!0.007$, $r\!<\!0.056$ \cite{Planck2018-X}.
\vspace {4pt}

\noindent{
\renewcommand\arraystretch{1.15}
\begin{tabular}{|c|c|c|c|c|c|c|c|}\hline
\multicolumn{8}{|c|}{Table 1}\\ \hline \hline
%
$p $&$q $&$\alpha $&$\beta$& $\varepsilon$ &$|\eta| $&$n_s$&$r=16\varepsilon$\\ \hline  \hline
$1$ & $1/2$ &$100$ &$1$ &$7.24\cdot 10^{-6}$ &$0.01507$ &$\,0.9698$ & $1.16\cdot 10^{-4}$ \\ \hline
$1$ & $1/2$ &$400$ &$2$ &$3.31\cdot 10^{-6}$ &$0.01699$ &$\,0.9660$ & $5.3\cdot 10^{-5}$ \\ \hline
$1$ & $1/2$ &$900$ &$3$ &$2.76\cdot 10^{-6}$ &$0.01747$ &$\,0.9650$ & $4.42\cdot 10^{-5}$ \\ \hline\hline
$3/4$ & $2/3$ &$40$ &$1$ &$4.27\cdot 10^{-6}$ &$0.01562$ &$\,0.9687$ & $6.83\cdot 10^{-5}$ \\ \hline 
$3/4$ & $2/3$ &$100$ &$2$ &$2.66\cdot 10^{-6}$ &$0.01711$ &$\,0.9658$ & $4.26\cdot 10^{-5}$ \\ \hline
$3/4$ & $2/3$ &$150$ &$3$ &$3.02\cdot 10^{-6}$ &$0.01743$ &$\,0.9651$ & $4.83\cdot 10^{-5}$ \\ \hline
\end{tabular}
\renewcommand\arraystretch{1.}
}
\vspace {4pt}

Now, let us consider the dynamics of the background inflaton field. In the
slow-roll regime this dynamics is described by Eq. (\ref{eq15}). For
potential (\ref{eq12}) it takes the form%
\begin{equation}
f_{\tau }=-\frac{2\sqrt{2}}{3}e^{-f}\left( f_{\chi }\right) ^{2}.\label{eq23}
\end{equation}%
Introducing the parameters $\delta =\chi ^{2p}/\alpha $, $\sigma =\beta
^{2}/\chi ^{2}$, we first consider the region of large $\chi $ where $\delta
\gg 1$, $\sigma \ll 1$. In this case%
\begin{equation}
f=\chi \left[ 1+q/\delta +O\left( 1/\delta ^{2},\sigma /\delta \right) \right] ,  \label{eq24}
\end{equation}%
and therefore%
\begin{equation}
f^{2p}/\alpha =\delta \left[ 1+2pq/\delta +O\left( 1/\delta ^{2},\sigma/\delta \right) \right] .  \label{eq25}
\end{equation}%
It follows that%
\begin{equation}
f_{\chi }=1-\frac{q\left( 2p-1\right) }{f^{2p}/\alpha }+O\left( 1/\delta^{2},\sigma /\delta \right) .  \label{eq26}
\end{equation}

Substituting (\ref{eq26}) into Eq. (\ref{eq23}) and integrating for large $f$, we obtain the asymptotic %
solution in implicit form%
\begin{equation}
P\left( f\right) =P\left( f_{ini}\right) -\frac{2\sqrt{2}}{3}\left( \tau -\tau _{ini}\right) ,  \label{eq27}
\end{equation}%
where%
\begin{equation}
P\left( f\right) \simeq e^{f}\left[ 1+2\alpha q\left( 2p-1\right) /f^{2p}%
\right] ,  \label{eq28}
\end{equation}%
and $f$ is given by Eq. (\ref{eq24}).When $\alpha =0$, this yields an exact
solution to Eq. (\ref{eq15}) for the Starobinsky potential.

Let us now consider the final phase of inflation when $\delta \ll \sigma
\lesssim 1$. Restricting ourselves to the case when $2pq=1$, from (\ref{eq13}%
) we find%
\begin{equation}
f=\frac{\alpha ^{q}}{\left( 1+\sigma \right) ^{1/2}}\left( 1+O\left( \delta\right) \right) ,  \label{eq29}
\end{equation}%
where $\delta =\chi ^{1/q}/\alpha $. It follows that in the leading order $%
\sigma \equiv \beta ^{2}/\chi ^{2}=\alpha ^{2q}/f^{2}-1$ and%
\begin{equation}
f_{\chi }=\frac{\sigma }{\delta ^{q}\left( 1+\sigma \right) ^{3/2}}=\frac{%
\alpha ^{q}}{\beta }\left( 1-\frac{f^{2}}{\alpha ^{2q}}\right) ^{3/2}.
\label{eq30}
\end{equation}%
Substituting this expresion into Eq. (\ref{eq23}) and introducing a new
variable%
\begin{equation}
x=f/\alpha ^{q}=\left( 1+\sigma \right) ^{-1/2}<1,  \label{eq31}
\end{equation}%
we obtain after integration the following solution in implicit form%
\begin{equation}
Q\left( x\right) =Q\left( x_{end}\right) +\frac{2\sqrt{2}\alpha ^{q}}{3\beta^{2}}\left( \tau _{end}-\tau \right) ,  \label{eq32}
\end{equation}%
where%
\begin{eqnarray}
Q\left( x\right) &=&\frac{1}{16}\left( \alpha ^{2q}+3\alpha ^{q}+3\right)
e^{-\alpha ^{q}}{\rm{Ei}}\left( \alpha ^{q}\left( x+1\right) \right) \nonumber \\
&&-\frac{1}{16}\left( \alpha ^{2q}-3\alpha ^{q}+3\right) e^{\alpha ^{q}}%
{\rm{Ei}}\left( \alpha ^{q}\left( x-1\right) \right)  \nonumber \\
&&+\frac{e^{\alpha ^{q}x}\left( 5x-\alpha ^{q}\left( 1-x^{2}\right)
-3x^{3}\right) }{8\left( 1-x^{2}\right) ^{2}}.  \label{eq33}
\end{eqnarray}

\subsection{Preheating stage}

At the preheating stage $\chi ^{2}\ll \beta ^{2}$, so that we can approximate%
\begin{equation}
f(\chi )\approx \chi \left( 1+\frac{\alpha }{\beta ^{2p}}\right) ^{q}=C\chi ,\label{eq34}
\end{equation}%
\begin{equation}
U(\chi )\approx \left( 1-e^{-C\chi }\right) ^{2}.  \label{eq35}
\end{equation}%
To describe the field oscillations at the minimum of this potential, it is
necessary to proceed from the complete system of equations (\ref{eq9})-(\ref{eq11}), %
considering it as a dissipative dynamical system with respect to
variables $\chi $ and $\chi _{\tau }$. Assuming $C\gg 1$, we use the technique %
of separation of fast (phase $\theta $) and slow (energy density $\varrho $) variables \cite{Moiseev}, setting%
\begin{equation}
\chi =\chi (\theta ,\varrho ),\quad \chi _{\tau }=\Omega (\varrho )\chi_{\theta }(\theta ,\varrho ),  \label{eq36}
\end{equation}%
where $\chi (\theta ,\varrho )$ is a $2\pi $-periodic in $\theta $ solution of the equation%
\begin{equation}
\Omega ^{2}(\varrho )\chi _{\theta \theta }+U_{\chi }=0,  \label{eq38}
\end{equation}%
and $\Omega (\varrho )\gg 1$ is a frequency. In the Van der Pol
approximation, the averaged equations for $\theta $ and $\varrho $ take the
form%
\begin{equation}
\theta _{\tau }=\Omega (\varrho ),\quad \varrho _{\tau }=-\left( 3/\sqrt{2}%
\right) \varrho ^{3/2}\gamma (\varrho ),  \label{eq39}
\end{equation}%
where%
\begin{equation}
\gamma (\varrho )=\left( 2/\varrho \right) \sqrt{1-\varrho }\left( 1-\sqrt{%
1-\varrho }\right)  \label{eq40}
\end{equation}%
(for details, see \cite{Koutvitsky2017, Koutvitsky2019}). Integrating Eqs. (\ref{eq38}) and (\ref{eq39}) gives:%
\begin{equation}
\Omega (\varrho )=C\sqrt{2\left( 1-\varrho (\tau )\right) },  \label{eq41}
\end{equation}%
\begin{equation}
\chi (\theta ,\varrho )=\frac{1}{C}\ln \frac{1-\varrho ^{1/2}(\tau )\cos\theta (\tau )}{1-\varrho (\tau )},  \label{eq42}
\end{equation}%
\begin{equation}
\theta(\tau)\!=\!\frac{2C}{3}\,\left[ g(\tau )\!-\!g(\tau _{0})\!-\! 2{\rm arctg\,}g(\tau )\!+\!2{\rm arctg\,}g(\tau _{0})\right],  \label{eq43}
\end{equation}
\begin{equation}
\varrho (\tau )=\frac{4g^{2}(\tau )}{\left( 1+g^{2}(\tau )\right) ^{2}}, \label{eq44}
\end{equation}%
where%
\begin{equation}
g(\tau )=\left( 3/\sqrt{2}\right) \left( \tau -\tau _{0}\right) 
+\varrho_{0}^{-1/2}\left( 1+\sqrt{1-\varrho _{0}}\right) ,  \label{eq45}
\end{equation}%
$\varrho _{0}=\varrho (\tau _{0})$, $\tau _{0}$ is the preheating stage start time.
 Note that $g(\tau )\geqslant 1$, since $\tau \geqslant\tau _{0}$ and $0<\varrho _{0}\leqslant 1$.

In addition, from Eqs. (\ref{eq10}) and (\ref{eq39}) we find %
\begin{equation}
\frac{a_{\varrho }}{a}=-\frac{1}{3\varrho\gamma (\rho )}\, ,  \label{eq46}
\end{equation}%
which gives%
\begin{equation}
\frac{a}{a_{0}}=\left( \frac{1-\sqrt{1-\varrho _{0}}}{1-\sqrt{1-\varrho(\tau )}}\right) ^{1/3}.  \label{eq47}
\end{equation}

\subsection{Approximate solutions vs numerical integration}

\begin{figure}[ht]
\begin{center}
\includegraphics[width=0.48\textwidth]{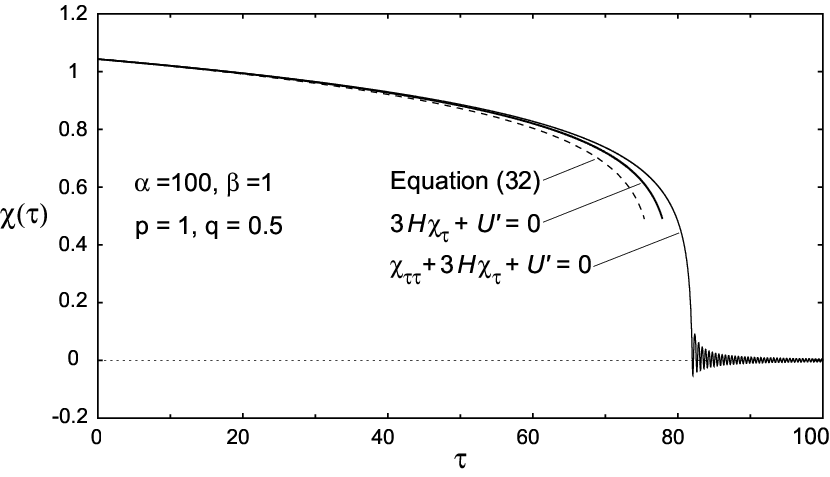}
\end{center}
\vspace{-18pt}
\caption{Numerical solutions of Eqs. (\ref{eq9}) and (\ref{eq15}), and approximate analytical solution (\ref{eq32})\label{Fig3}}
\end{figure}

To verify our results, we numerically integrated Eq. (\ref{eq9}), as well as
Eq. (\ref{eq15}), with the initial conditions%
\[
\chi (0)=\chi _{\ast },\quad \chi _{\tau }(0)=0, 
\]%
where $\chi _{\ast }$ was found from Eq. (\ref{eq22}) with $N_{sr}=55$. The
same value of $\chi _{\ast }$ was used to calculate initial value of $x$ in
analytical solution (\ref{eq32}). The results of the comparison are shown in
Fig \ref{Fig3}. 

\begin{figure}[ht]
\begin{center}
\includegraphics[width=0.47\textwidth]{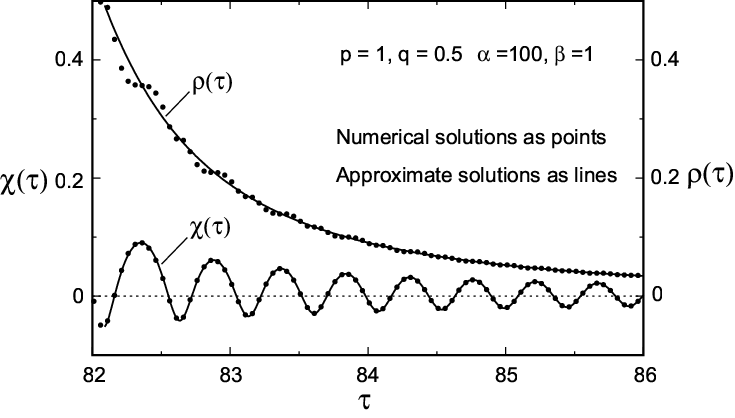}
\end{center}
\vspace{-16pt}
\caption{Analytical (solid lines) and numerical (points) solutions, describing the preheating stage\label{Fig4}}
\end{figure}

To compare our approximate solutions (\ref{eq41})-(\ref{eq45}), describing the preheating, %
with the results of numerical integration, we set the initial condition%
$$
\chi (\theta (\tau _{0}),\varrho (\tau _{0}))=\chi _{0}, 
$$
where $\tau _{0}$ is the moment, when the field $\chi (\tau )$, obtained by
numerical integration of Eq. (\ref{eq9}), reaches its first minimum at the
begining of oscillations, so that $\chi _0\!=\!\chi (\tau_0)\!<0$. This gives
the initial value%
$$
\varrho _{0}=\left( 1-e^{-C \chi_{0}}\right) ^{2}, 	
$$
involving in the function $g(\tau )$ (\ref{eq45}). As can be seen from the
graphs in Fig. \ref{Fig4}, the obtained solutions practically coincide even 
over a fairly large time interval.

\section{Perturbations: resonant phenomena during preheating}

As was found above, at the preheating stage the spatially homogeneous part
of the inflaton field, $\chi \left( \tau \right) =\chi \left( \theta
,\varrho \right) $, coherently oscillates at the minimum of the potential (%
\ref{eq35}) in accordance with Eqs. (\ref{eq42})-(\ref{eq45}). Now let us
consider the fluctuations of the inflaton field above this oscillating
background and the associated perturbations of the metric. The key point in
our consideration will be the study of resonance effects in this system.
When doing this in a linear approximation, it is sufficient to consider 
the evolution of only individual perturbation modes with standard  
initial conditions used in Floquet theory.

\subsection{Perturbations of the inflaton field}

Let us first consider perturbations $\delta \chi \left( \tau ,\mathbf{x}\right) $ %
of the inflaton field, neglecting perturbations of the metric. In
the Friedman-Robertson-Walker background metric, the spatial modes of these
perturbations satisfy the equation%
\begin{equation}
\left( \delta \chi _{k}\right) _{\tau \tau }+3\mathcal{H}\left( \delta \chi
_{k}\right) _{\tau }+\left( q^{2}+U_{\chi \chi }\right) \delta \chi _{k}=0,
\label{eq48}
\end{equation}%
where $q=b^{-1}V_{\infty }^{-1/2}\left( k/a\right) $ is the normalized
physical momentum. Setting%
\begin{equation}
\delta \chi _{k}=\left( a(\rho)/a_{0}\right) ^{-3/2}\Omega^{-1/2}\left( \rho \right) X_{k}\left( \theta \right)  \label{eq49}
\end{equation}%
with $\varrho =\varrho \left( \theta \right) $ and using Eqs. (\ref{eq10}), (\ref{eq35}), and (\ref{eq39})-(\ref{eq42}), %
we arrive at the equation%
\begin{equation}
\frac{d^{2}X_{k}}{d\theta ^{2}}+\left( \frac{q^{2}}{\Omega ^{2}}+%
\frac{1-2\varrho +\varrho ^{1/2}\cos \theta }{\left( 1-\varrho ^{1/2}\cos \theta\right) ^{2}}\right) X_{k}=0,  \label{eq50}
\end{equation}%
where the phase $\theta $ is used as a new time variable instead of $\tau $,
and the terms of the orders $\lesssim \varrho /\Omega ^{2}\sim \left( \mathcal{H}/\Omega \right) ^{2}\ll 1$ %
are neglected in the brackets.

Since%
\begin{equation}
\frac{d}{d\theta }\ln q=-\frac{\mathcal{H}}{\Omega },  \label{eq51}
\end{equation}%
\begin{equation}
\frac{d\varrho }{d\theta }=-6\frac{\mathcal{H}}{\Omega }\sqrt{1-\varrho }\left( 1-\sqrt{1-\varrho }\right) ,  \label{eq52}
\end{equation}%
Eq. (\ref{eq50}) can be regarded as the Hill equation with slowly varying
parameters $q$ and $\varrho $ related by%
\begin{equation}
q=q_{0}\left( \frac{1-\sqrt{1-\varrho }}{1-\sqrt{1-\varrho _{0}}}\right)^{1/3}.  \label{eq53}
\end{equation}%
For $\varrho \ll 1$, Eq. (\ref{eq50}) takes the form of the Mathieu equation%
\begin{equation}
\frac{d^{2}X_{k}}{d\theta ^{2}}+\left( \frac{q^{2}}{\Omega ^{2}}+1+3\varrho^{1/2}\cos \theta +O\left( \varrho \right) \right) X_{k}=0.  %
\label{eq53a}
\end{equation}

\begin{figure}[ht]
\begin{center}
\includegraphics[width=0.47\textwidth]{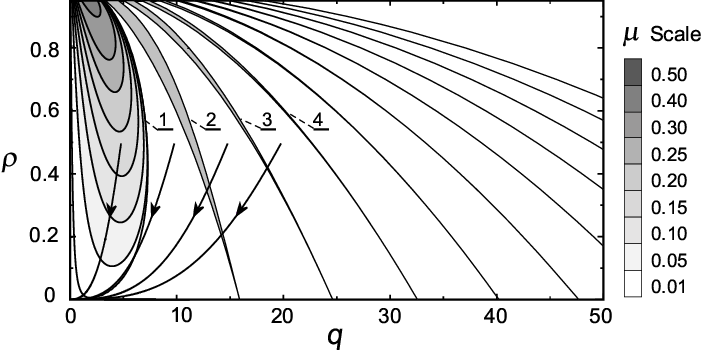}
\end{center}
\vspace{-18pt}
\caption{Resonance zones of Eq. (\ref{eq50}) for $C^2=101$). %
$\mu$-Scale relates to the broad zone \label{Fig5}}
\end{figure}

Neglecting the cosmological expansion, the parameters $q$ and $\varrho $
become constants. Fig. \ref{Fig5} shows the parametric resonance zones obtained by
numerical integration of Eq. (\ref{eq50}) with fixed values of $q$ and $\varrho $. 
In this case, the growing solutions in the resonance zones have the form %
$X_{k}(\theta) \sim A_{k}(\theta) e^{\mu (q,\varrho) \theta}$, 
where $A_{k}(\theta)$ is a periodic function of $\theta$, and $\mu(q,\varrho)$ is the Floquet exponent. 

\begin{figure}[ht]
\begin{center}
\includegraphics[width=0.45\textwidth]{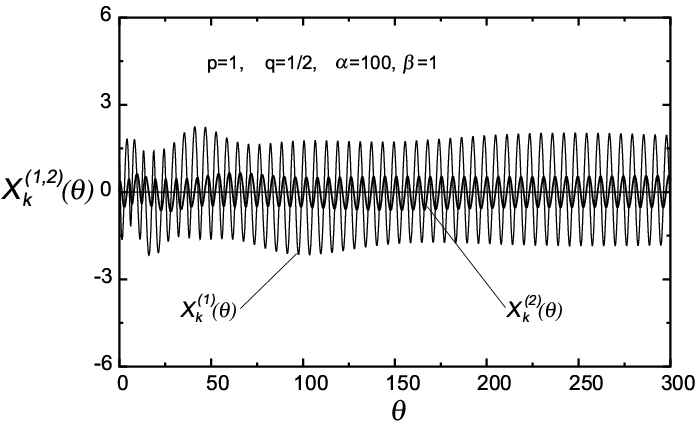}
\end{center}
\caption{Normalized solutions $X_{k}^{(1)}(\theta) $ and $X_{k}^{(2)}(\theta) $ %
along the trajectory with $q_0=15,$ $\varrho_0=0.497$, passing through zone 2 in Fig. \ref{Fig5}\label{Fig6}}
\end{figure}

\begin{figure}[ht]
\begin{center}
\includegraphics[width=0.45\textwidth]{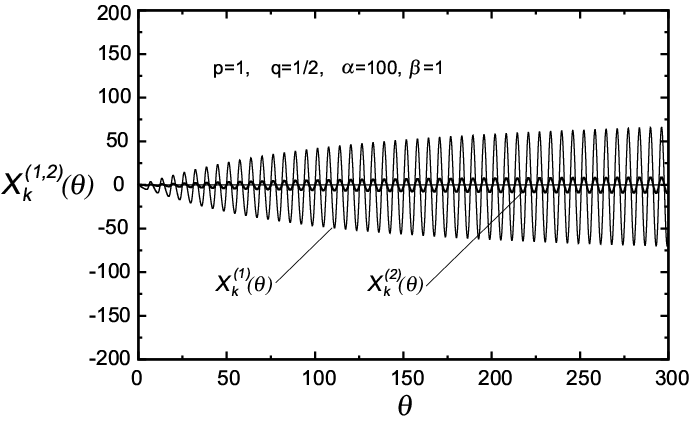}
\end{center}
\caption{Normalized solutions $X_{k}^{(1)}(\theta) $ and $X_{k}^{(2)}(\theta) $ %
along the trajectory with $q_{0}=5,$ $\varrho_{0}=0.497$, passing through zone 1 in Fig. \ref{Fig5}\label{Fig7}}
\end{figure}

Taking into account the cosmological expansion, the point $\left( q,\varrho
\right) $ with initial coordinates $\left( q_{0},\varrho _{0}\right) $
slowly moves along the trajectory (\ref{eq53}) in accordance with Eqs. (\ref%
{eq51}) and (\ref{eq52}). In this case, when passing through a resonant
zone, the growing solution we are interested in takes the form $X_{k}\left(
\theta \right) \sim A_{k}\left( \theta \right) \exp \int \mu \left(
q,\varrho \right) d\theta $ \cite{Karsten_Jedamzik_2010, Koutvitsky2018}. This solution can be represented
as%
\begin{equation}
X_{k}\left( \theta \right) =c_{1}X_{k}^{(1)}\left( \theta \right)
+c_{2}X_{k}^{(2)}\left( \theta \right) ,  \label{eq53e}
\end{equation}%
where $X_{k}^{(1)}\left( \theta \right) $ and $X_{k}^{(2)}\left( \theta
\right) $ are independent normalized solutions of Eq. (\ref{eq50}) with
initial conditions%
\begin{eqnarray}
\,X_{k}^{(1)} &=&1,\;dX_{k}^{(1)}/d\theta =0,  \nonumber \\
X_{k}^{(2)} &=&0,\;dX_{k}^{(2)}/d\theta =1,  \label{eq53c}
\end{eqnarray}%
at $\theta =0$, and the constants $c_{1,2}$ satisfy the ratio%
\begin{equation}
\frac{c_{2}}{c_{1}}=\frac{e^{2\pi \mu }-X_{k}^{(1)}\left( 2\pi \right) }{%
X_{k}^{(2)}\left( 2\pi \right) },  \label{eq53d}
\end{equation}%
calculated at the starting point $\left( q_{0},\varrho _{0}\right) $.
Therefore, to study resonance effects at the preheating stage, it will be
sufficient for us to consider the behavior of normalized solutions $%
X_{k}^{(1)}$ and $X_{k}^{(2)}$, as is usually done in Floquet theory (see
e.g., \cite{Magnus+Winkler}). Examples of these solutions, obtained by numerical
integration of Eq. (\ref{eq50}), are shown in Figs. \ref{Fig6} and \ref{Fig7}.

As can be seen from %
Fig. \ref{Fig6}, during the passage of the narrow resonant zone
2, the amplitude of the oscillations remains practically unchanged (of the
order of unity). Indeed, for sufficiently small $\varrho $, zone 2 in Fig. \ref{Fig5}
corresponds to the third resonant zone of the Mathieu equation (\ref{eq53a}%
). The width of this zone can be estimated as $\Delta q\lesssim \Omega
\varrho ^{3/2}$ \cite{Abramowitz+Stegun}. Taking into account Eq. (\ref{eq51}), we find
that the time required for the trajectory to pass through zone 2 near $%
q\simeq \Omega $ is estimated as $\Delta \theta \lesssim \left( \Omega /%
\mathcal{H}\right) \varrho ^{3/2}$. For example, for $\varrho \lesssim \pi
/C $, this means that $\Delta \theta $ is less than the period of the
background field oscillations. Therefore, in this case, parametric resonance
does not have time to develop, as can be seen in Fig. \ref{Fig6}. 
Furthermore, we note that the approximation based on the Mathieu equation (\ref{eq53a}) %
is rather crude, since it does not take into account the anharmonicity of the	%
background field oscillations. Therefore, the above estimates are somewhat
overestimated.

\begin{figure}[ht]
\begin{center}
\includegraphics[width=0.44\textwidth]{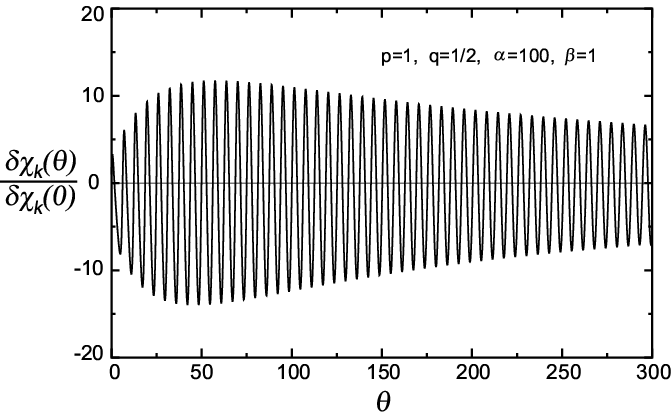}
\end{center}
\vspace{-8pt}
\caption{Evolution of $\delta\chi_{k}(\theta)$ along the trajectory in zone 1, %
obtained using Eq. (\ref{eq49}), where $X_{k}(\theta)=c_1 X_{k}^{(1)}(\theta) +c_2 X_{k}^{(2)}(\theta)$, %
$X_{k}^{(1,2)}(\theta)$ are normalized solutions shown in Fig. \ref{Fig7}, $c_{2}/c_{1}=8.6025$\label{Fig8}}
\end{figure}

In contrast, when passing through the broad zone, the oscillations initially
increase significantly (Fig. \ref{Fig7}). However, the increase in the amplitude of
these oscillations gradually slows down since $\mu \left( q,\varrho \right) $
approaches zero along the trajectory (see Fig. \ref{Fig5}), and consequently, $\int
\mu \left( q,\varrho \right) d\theta $ approaches a constant. As a result,
increasing the scale factor $a\left( \rho \right) $ and the background field
oscillation frequency $\Omega \left( \rho \right) $ included in Eq. (\ref%
{eq49}) leads to a gradual decrease in the amplitude of the inflaton field
perturbations, as seen in Fig. \ref{Fig8}. We found that at large times this amplitude
decreases approximately  as $\theta^{-1} $.

\subsection{Perturbations of the inflaton field and scalar perturbations of the metric}

The scalar modes of the perturbations of the inflaton field and the metric, $%
\delta \chi _{k}$ and $\Phi _{k}$, are described by the Mukhanov-Sasaki
equation \cite{PismaZhETF.41.402, PTP.76.1036} (see, e.g., \cite{Gorbunov} for a review)%
\begin{equation}
u_{k}^{\prime \prime }+\left( k^{2}-\frac{z^{\prime \prime }}{z}\right)
u_{k}=0,  \label{eq54}
\end{equation}%
where $\left( ^{\prime }\right) =d/d\eta =a\,d/dt=abV_{\infty }^{1/2}d/d\tau 
$, $\eta $ is conformal time, 
\begin{equation}
u_{k}=\frac{a_{0}}{b}\bar{u}_{k},\quad z=\frac{a_{0}}{b}\bar{z},
\label{eq55}
\end{equation}%
and the normalized functions $\bar{u}_{k}$ and $\bar{z}$ are given by 
\begin{equation}
\bar{u}_{k}=\left( a/a_{0}\right) \,\delta \chi _{k}+\bar{z}\Phi _{k},
\label{eq56}
\end{equation}%
\begin{equation}
\bar{z}=a\chi _{\tau }/\left( a_{0}\mathcal{H}\right) .  \label{eq57}
\end{equation}%

\begin{figure}[hb]
\parbox[t][0.3\textwidth][t]{0.47\textwidth}
{
\begin{center}
\includegraphics[width=0.46\textwidth]{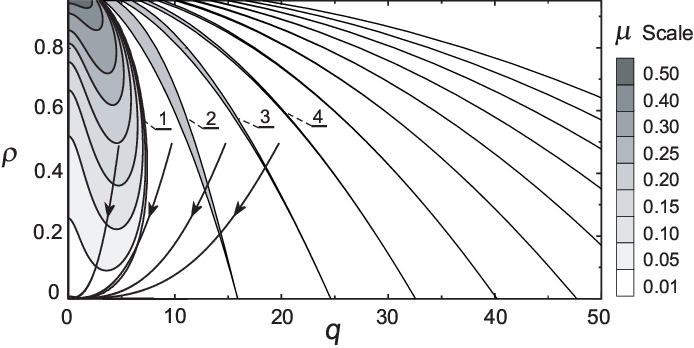}
\end{center}
\caption{Resonance zones of Eq. (\ref{eq59}) for $C^2=101$. $\mu$-Scale relates to the broad zone \label{Fig9}}
}
\end{figure}

\begin{figure}[ht]
{
\begin{center}
\includegraphics[width=0.45\textwidth]{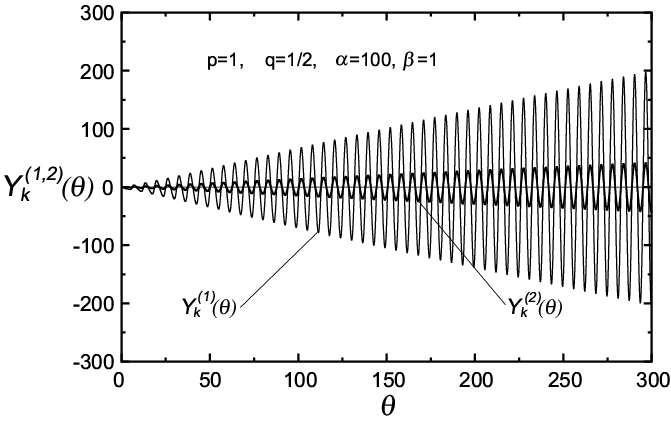}
\vspace{-12pt}
\end{center}
\caption{Normalized solutions $Y_{k}^{(1)}(\theta)$ and $Y_{k}^{(2)}(\theta)$ %
along the trajectory with $q_{0}=5,$ $\varrho_{0}=0.497$, passing through the broad zone in Fig. \ref{Fig9}\label{Fig10}}
}
\end{figure}

\noindent In terms of cosmic time $\tau $ and physical momentum $q$ Eq. (\ref{eq54})
takes the form%
\begin{eqnarray}
\frac{d^{2}F_{k}}{d\tau ^{2}}+&\Bigg[&q^{2}+\frac{d^{2}U}{d\chi ^{2}}+\frac{3%
\sqrt{2}}{\varrho ^{1/2}}\frac{d\chi }{d\tau }\frac{dU}{d\chi }-\frac{9}{8}\,U\nonumber\\
&+&\frac{81}{16}\left( \frac{d\chi }{d\tau }\right) ^{2}-\frac{9}{4\varrho }%
\left( \frac{d\chi }{d\tau }\right) ^{4}\Bigg] F_{k}=0,  \label{eq57a}
\end{eqnarray}
where $F_{k}=\left( a/a_{0}\right) ^{1/2}\bar{u}_{k}$ \cite{PhysRevLett.82.1362}. Setting%
\begin{equation}
\bar{u}_{k}=\left( a\left( \varrho \right) /a_{0}\right) ^{-1/2}\Omega
^{-1/2}\left( \varrho \right) Y_{k}\left( \theta \right) ,  \label{eq58}
\end{equation}%
with $\varrho =\varrho(\theta) $ and using Eqs. (\ref{eq2})-(\ref{eq4}), %
(\ref{eq9})-(\ref{eq11}), (\ref{eq35}), and (\ref{eq39})-(\ref{eq42}), %
from Eq. (\ref{eq57a}) we obtain the Hill equation of the form%
\begin{eqnarray}
\hspace{-10pt}\frac{d^{2}Y_{k}}{d\theta ^{2}}&+&\Bigg(\frac{q^2}{\Omega^2}+\frac{1-2\varrho +\varrho^{1/2}\cos\theta}{( 1-\varrho^{1/2}\cos\theta)^2}\nonumber\\
&+&\frac{6\varrho ^{1/2}}{C}\sqrt{1-\varrho }\,\frac{( \varrho^{1/2}\!-\cos \theta ) \sin \theta }{( 1-\varrho ^{1/2}\cos \theta) ^{3}}\Bigg) Y_{k}=0,  \label{eq59}
\end{eqnarray}
where, as before, the slowly varying parameters $q$ and $\varrho $ are
related by Eq. (\ref{eq53}), and the terms of the orders $\lesssim \varrho
/\Omega ^{2}\sim \left( \mathcal{H}/\Omega \right) ^{2}\ll 1$ are neglected.
The third term in the brackets describes the effect of metric perturbations
(see Eq. (\ref{eq50}) for comparison). Fig. \ref{Fig9} shows the parametric resonance
zones obtained by numerical integration of Eq. (\ref{eq59}) for fixed $q$
and $\varrho $. It can be seen that in region $q^{2}\gg C\varrho ^{1/2}$,
the narrow resonant zones are practically identical to those shown in Fig. \ref{Fig5}, 
since in this region the third term in Eq. (\ref{eq59}) can be neglected.
However, in region $q^{2}\lesssim C\varrho ^{1/2}$, this term leads to a
significant change in the structure of the broad zone, resulting in a sharp
increase in the oscillation amplitude due to parametric resonance (see Fig. \ref{Fig10}).

Now let us find $\Phi _{k}$. We start from the Poisson equation for the
gravitational potential $\Phi $ (see, e.g., \cite{Gorbunov}):%
\begin{equation}
\Delta\Phi=\frac{3a_{0}}{4a}\bar{z}\chi^{\prime }\left( \frac{\bar{u}}{\bar{z}}\right)^{\prime }.  \label{eq60}
\end{equation}%
In terms of Fourier components and cosmic time this equation can be written
as%
\begin{equation}
q^{2}\Phi _{k}=-\frac{3}{4}\mathcal{H}\left( \frac{a_{0}}{a}\right)
^{2}\left( \bar{z}\frac{d\bar{u}_{k}}{d\tau }-\bar{u}_{k}\frac{d\bar{z}}{%
d\tau }\right) .  \label{eq61}
\end{equation}%
Using Eqs.(\ref{eq39})-(\ref{eq42}), and (\ref{eq47}), we obtain%
\begin{eqnarray}
\Phi _{k}\simeq &-&\frac{3}{4q^{2}}\Omega ^{1/2}\left( \frac{a}{a_{0}}\right)
^{-3/2}\frac{\sqrt{2\varrho \left( 1-\varrho \right) }}{1-\varrho ^{1/2}\cos
\theta }\times\nonumber\\
&&\left( \frac{\varrho ^{1/2}-\cos \theta }{1-\varrho ^{1/2}\cos
\theta }Y_{k}+\sin \theta \,\frac{dY_{k}}{d\theta }\right) ,  \label{eq62}
\end{eqnarray}%
where the terms $\sim \left( \varrho ^{1/2}/C\right) \left(
Y_{k},\,dY_{k}/d\theta \right) $ in the brackets are neglected. As a result,
substituting this expression into formula (\ref{eq56}) with the same
accuracy gives%
\begin{eqnarray}
&&\delta \chi _{k}=\frac{a_{0}}{a}\left( \bar{u}_{k}-\bar{z}\Phi _{k}\right) \simeq \nonumber\\
\Omega ^{-1/2}\left( \frac{a}{a_{0}}\right) ^{-3/2}&&\Bigg[ Y_{k}\left(\theta \right) +\frac{3C\varrho ^{1/2}\left( 1\!-\!\varrho \right) ^{3/2}\sin\theta }{q^{2}\left( 1-\varrho ^{1/2}\cos \theta \right) ^{2}}\times\nonumber\\
&&\hspace{-24pt}\left( \frac{\varrho ^{1/2}-\cos \theta }{1-\varrho ^{1/2}\cos \theta }\,Y_{k}\left(\theta \right) +\sin \theta \,\frac{dY_{k}}{d\theta }\right) \Bigg].
\label{eq63}
\end{eqnarray}%

It is clear that when $q^{2}\gg C\varrho ^{1/2}$ we return to Eqs. (\ref{eq49}) and %
(\ref{eq50}). Fig. 11 shows example of the evolution of scalar
modes of the metric and inflaton field perturbations in the broad resonance
zone. It is evident that cosmological expansion significantly suppresses the
effects of parametric resonance, limiting the growth of the oscillation
amplitude.
\begin{figure}[ht]
\begin{center}
\includegraphics[width=0.45\textwidth]{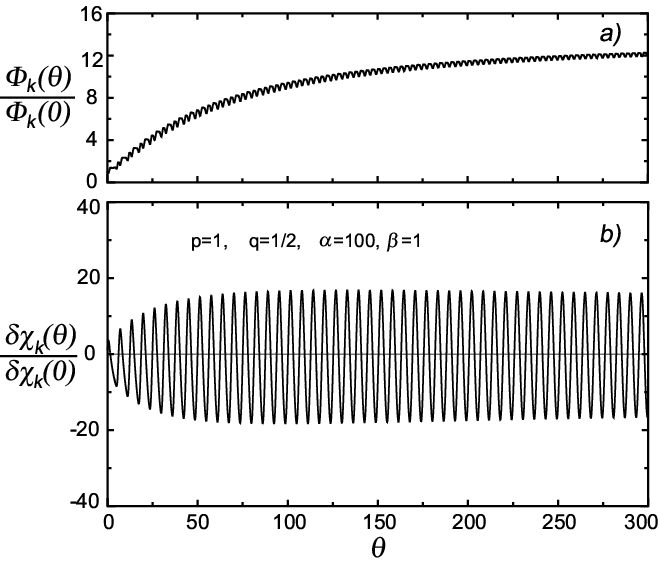}
\end{center}
\vspace{-12pt}
\caption{Evolution of the spatial modes of the gravitational potential $\Phi_{k}$ (a) %
and inflaton field perturbation $ \delta\chi_{k}$ (b) along the trajectory in zone 1,
obtained using Eqs. (\ref{eq62}) and (\ref{eq63}), where %
$Y_{k}(\theta)=c_1 Y_{k}^{(1)}(\theta) +c_2 Y_{k}^{(2)}(\theta) $,\ 
$Y_{k}^{(1,2)}(\theta)$ are normalized solutions shown in Fig. \ref{Fig10}, 
$c_2/c_1=12.050$ (see Eqs. (\ref{eq53c}) and (\ref{eq53d})) \label{Fig11}}
\end{figure}

\subsection{Tensor perturbations of the metric}

\begin{figure}[ht]
\begin{center}
\includegraphics[width=0.45\textwidth]{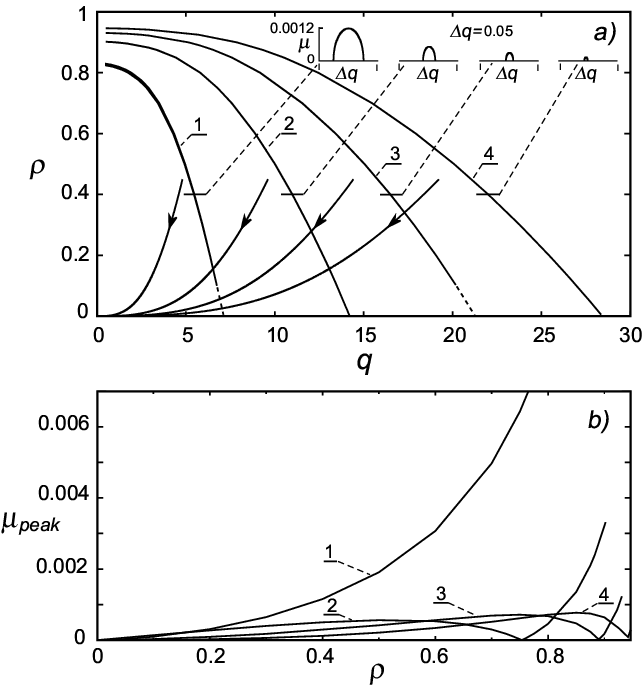}
\end{center}
\caption{The first few resonant zones of Eq. (\ref{eq68}) for $C^2=101$ (a) %
and peak values of the Floquet exponent along them (b)}\label{Fig12}
\end{figure}

The Fourier components $h_{k}$ of the tensor perturbations are described by
the equation \cite{Grishchuk1975}%
\begin{equation}
v_{k}^{\prime \prime }+\left( k^{2}-\frac{a^{\prime \prime }}{a}\right)v_{k}=0,  \label{eq64}
\end{equation}%
where $v_{k}=\left( a/a_{0}\right) h_{k}$. In terms of the cosmic time $\tau$ %
and physical momentum $q$ this equation takes the form (see e.g., \cite{Karsten_Jedamzik_2010}) 
\begin{equation}
\frac{d^{2}G_{k}}{d\tau ^{2}}+\left[ q^{2}+\frac{9}{8}\left( \frac{\chi
_{\tau }^{2}}{2}-U\left( \chi \right) \right) \right] G_{k}=0,  \label{eq65}
\end{equation}%
where $G_{k}=\left( a/a_{0}\right) ^{3/2}h_{k}$. Notice, that 
\begin{equation}
\chi _{\tau }^{2}/2-U\left( \chi \right) =\varrho -2U\left( \chi \right) 
\label{eq66}
\end{equation}%
is the normalized background pressure. Taking this into account and setting%
\begin{equation}
h_{k}=\left( a\left( \rho \right) /a_{0}\right) ^{-3/2}\Omega ^{-1/2}\left(
\rho \right) Z_{k}\left( \theta \right) ,  \label{eq67}
\end{equation}%
from Eq. (\ref{eq65}) we obtain the Hill equation of the form%

\begin{eqnarray}\label{eq68}
\frac{d^{2}Z_{k}}{d\theta ^{2}}&+&\frac{1}{\Omega ^{2}}\Bigg\{ q^{2}+\frac{9}{8%
}\varrho \Bigg[ 1-2\left( \frac{\varrho ^{1/2}-\cos \theta }{1-\varrho
^{1/2}\cos \theta }\right) ^{2}\nonumber\\
&+&\frac{\varrho \gamma \left( \varrho \right)}{1-\varrho }\left( 1+\frac{\left( 2+\varrho \right) \gamma \left( \varrho
\right) }{4\left( 1-\varrho \right) }\right) \Bigg] \Bigg\} Z_{k}=0.
\end{eqnarray}
Fig. \ref{Fig12} shows the parametric resonance zones obtained numerically by
integrating this equation for fixed $q$ and $\varrho $. It can be seen that
these zones are very narrow and have extremely small values of the Floquet
exponent. As $\varrho $ approaches zero, zones 1 and 3 disappear. 
The remaining zones 2 and 4 in this limit correspond to the first and second
resonant zones of the Mathieu equation %
\begin{equation}
\frac{d^{2}Z_{k}}{d\theta ^{2}}+\frac{1}{\Omega ^{2}}\left( q^{2}-\frac{9}{8}%
\varrho \cos 2\theta +O\left( \varrho ^{3/2}\right) \right) Z_{k}=0,
\label{eq69}
\end{equation}%
which is obtained from Eq. (\ref{eq68}) for $\varrho \ll 1$. We use this
fact to estimate the possibility of amplification of gravitational
fluctuations when passing through the resonant zones of Eq. (\ref{eq68}).
Let us consider, for example, zone 2 of this equation. For sufficiently
small $\varrho $, the width of the corresponding zone of the Mathieu
equation (\ref{eq69}) in the neighborhood of $q\simeq \Omega $ is estimated as 
$\Delta q\lesssim \varrho /\Omega $. Therefore, in view of Eq. (\ref{eq51}),
the characteristic time of passage through this resonant zone will be $%
\Delta \theta \lesssim \varrho ^{1/2}/\Omega \ll T=2\pi $. Thus, we conclude
that resonant amplification of gravitational fluctuations does not occur,
which is in agreement with the result of Ref. \cite{Karsten_Jedamzik_2010}. Moreover, in
the limit $\varrho \rightarrow 0$, the solution of Eq. (\ref{eq69}) can be
obtained in the WKB approximation, in which $Z_{\kappa }\propto \omega
^{-1/2}$, where $\omega =q/\Omega $ is the effective frequency. From (\ref%
{eq67}) it then follows that as the Universe expands, the amplitude of the
tensor modes decreases as $a^{-1}$.

\section{Concluding remarks}

In this paper, we have investigated the dynamics of the inflaton scalar
field $\chi $ in a four-parameter potential that behaves like the
Starobinsky potential for large values of $\chi $ and like the $\alpha $%
-attractor potential for small values of $\chi $ (see Eqs. (\ref{eq12}) and (%
\ref{eq13})). Using the slow-roll approximation, we have found the analytic
expressions (\ref{eq27}), (\ref{eq28}) and (\ref{eq32}), (\ref{eq33})
describing evolution of the background inflaton field in this potential at
the inflationary stage. When considering the preheating stage, we assumed that
the constant $C$ in Eq. (\ref{eq34}) is large enough. This means that the
potential $U\left( \chi \right) $ has a narrow well near zero, in which
rapid, low-amplitude oscillations of the uniform background field occur.
However, these oscillations are nonlinear, especially at the beginning of
preheating. For this reason, we did not use a parabolic approximation of the
potential and did not even expand it in powers of $\chi $ to study the
effect of anharmonicity, as is usually done (see, e.g., \cite{Karsten_Jedamzik_2010,
Richard_Easther_2011, Martin_2020, PhysRevD.108.063524, DELCORRAL2024169824}). 
Instead, to describe the background oscillations we used the method of separating the fast and slow variables 
\cite{Koutvitsky2017, Koutvitsky2019}, assuming that the oscillation frequency $\Omega $ is
much greater than the expansion rate of the Universe, i.e, $\mathcal{H}%
/\Omega \ll 1$. As a result, we have obtained expressions (\ref{eq41})-(\ref%
{eq45}), and (\ref{eq47}), describing the damped background field oscillations
with the fast phase $\theta $, as well as the slow (compared to $\theta $)
evolution of the energy density $\varrho $ and the scale factor $a$. We used
these expressions in studying perturbations of the inflaton and the metric.
Based on the Mukhanov-Sasaki equation we have derived the Hill equation (\ref%
{eq59}) with slowly varying parameters, which describes the scalar
perturbation modes. The corresponding Hill equation has been obtained for
tensor modes.

It should be noted that these equations, as well as the Mukhanov-Sasaki
equation, were derived in the linear approximation in $\delta \chi $, $\Phi $%
, and $h$. In particular, this means that $\left\vert \delta \chi
\right\vert \ll \left\vert \chi \right\vert $. But since $\left\vert \chi
\right\vert $ decays as $\varrho ^{1/2}/C$ while the amplitude of $\delta
\chi $ remains practically constant (see Eq. (\ref{eq42}) and Fig. 11), this
condition will be violated at late times. As follows from results of lattice
simulations of $\alpha $-attractor model (\ref{eq6}) performed in Refs. \cite{PhysRevD.108.063524, DELCORRAL2024169824}, %
for values of $\alpha \lesssim 10^{-4}$ the nonlinear
regime, when $\left\vert \delta \chi \right\vert \gtrsim \left\vert \chi
\right\vert $, occurs very quickly, after several background oscillations.
This can lead to a breakdown in the coherence of oscillations due to the
backreaction effect and the formation of localized nonlinear structures such
as oscillons and gravipulsons \cite{PhysRevD.108.063524, PhysRevLett.108.241302, PhysRevD.83.124028}. %
Taking this into account, we assume that our results regarding the preheating stage are valid
for $1\ll C\lesssim 10^{3/2}$, where the upper bound corresponds to the
estimate obtained in Ref. \cite{DELCORRAL2024169824} with consideration for metric
fluctuations. We leave the study of the nonlinear structures arising during
the preheating for a future work.

\section*{Acknowlegements}

We would like to thank E.~A. Kuznetsov and the participants of the XXXIV session of the 
RAS Council on Nonlinear Dynamics for helpful discussions. 
This work was supported by the scientific assignment of  IZMIRAN.

\bibliography{maslov.bib}

\end{document}